\documentstyle[LTpaper,epsfig]{article}
\begin{document}
%
%
\title{Transverse optical plasmons in layered superconductors.}
%
%
\author{D. van der Marel$^a$,and A. Tsvetkov$^{a,b}$}

%
\address{$^a$Materials Science Center, Laboratory of Solid State Physics,
University of Groningen,\\
Nijenborgh 4, 9747 AG Groningen, The Netherlands\\
$^b$P. N. Lebedev Physical Institute, Russian Academy of Sciences, 
Leninsky prospect 53, 117924 Moscow, Russia}
%
\abstract{We discuss the possible existance of transverse optical
plasma modes in superlattices consisting of Josephson
coupled superconducting layers. These modes appear as resonances
in the current-current correlation function, as opposed to the
usual plasmons which are poles in the density-density channel. 
We consider both bilayer superlattices, and single layer lattices
with a spread of interlayer Josephson couplings. We
show that our model is in quantitative agreement with
the recent experimental observation by a number of groups of 
a peak at the Josephson plasma frequency 
in the optical conductivity of La$_{1.85}$Sr$_{0.15}$CuO$_4$.}
\maketitle\pagestyle{empty}

%
%
\section{INTRODUCTION}
In recent years interlayer Josephson plasmons have received considerable
attention both theoretically and experimentally. So far this attention has
been restricted to systems consisting of a stack of identical 2-dimensional
superconducting layers coupled by identical insulating barriers. In this
paper we consider the more general problem of ordered and disordered
superlattice structures.
Regardless of the underlying microscopic model, in the superconducting
state the low energy scale properties of a multilayer of Josephson
coupled 2-dimensional superconducting layers is described by the
Lawrence-Doniach model\cite{lawrence}. This model has been applied
to the case of an ordered array of single superconducting layers,
by Bulaevskii {\em et al.}\cite{bulaevskii1}, who calculated the 
longitudinal dielectric function, and by Tachiki {\em et al.}\cite{tachiki},
who studied the transverse dielectric function. Microscopically the
interlayer Josephson resonances are a result of broken gauge symmetry in
a superconductor with long range Coulomb interactions\cite{anderson}, and
belong to a larger class of several types of
collective modes in layered unconventional 
superconductors\cite{modes}. A low frequency Josephson plasmon\cite{uchida}
arises as a direct consequence of the electrodynamical properties of a 
BCS superconductor in the dirty limit\cite{kim,pokrovsky,zha}.
A different scenario has been discussed by Anderson, in which these modes 
occur as a direct consequence of the
interlayer pairing mechanism for superconductivity in the high 
T$_c$ materials\cite{chakravarty,phil}. In this spirit we 
attempt to set the stage for further experimental studies of
layered superconductors, in particular materials with more
than one superconducting layer per unit cell 
({\em e.g.} Bi$_2$Sr$_2$CaCu$_2$O$_8$)
and materials where the interlayer Josephson parameters are 
randomly distributed due to disorder ({\em e.g.} 
La$_{2-x}$Sr$_{x}$CuO$_{4}$\cite{uchida,kim}). 
\section{RESULTS}
To simplify the discussion we will only consider the charge 
dynamics for $\vec{k}\rightarrow 0$, with the electric field 
vector perpendicular to the CuO$_2$ planes, which is the relevant limit 
for a discussion of the $c$-axis optical properties. A discussion of the
dielectric tensor for general values of $k$ was given by Bulaevskii 
for the superconducting state\cite{bulaevskii1} and by one of us for 
the normal state\cite{stanfordproc}. Let us consider the relevant
Hamiltonian of a stack of superconducting planes coupled through
the Josephson effect, each with a total area $A$, and with total thickness
$D$ of the crystal along the $z$ direction.
\\
Each plane is characterized by a charge per unit area $\rho_m$ 
and a phase $\phi_m$, where $m$ is the layer index. 
As we are only considering electrical fields perpendicular to the
planes, the long-range Coulomb potential between each set of planes
increases linearly with distance. The 
charge dynamics enters via the Josephson coupling $J_m$ between 
each set of nearest neighbor planes. 
The equations of motion for a general distribution of interlayer
couplings, and general values of wavevector $\vec{k}$ were derived by 
Bulaevskii\cite{bulaevskii1}. In the present paper we only consider
$\vec{k}\rightarrow 0$, relevant for optical experiments. 
The solution for general values of $\vec{k}$ will be presented 
elsewhere\cite{general}.
The limit $\vec{k}\rightarrow 0$ leads to a considerable simplification of the
equations of motion, which in this case correspond exactly to an equivalent
network model where each set of two neighbouring layers can be represented 
by a capacitance shunted by an induction. The total complex impedance $Z$ 
of the crystal is the series resistance of all local impedances
$ Z_m  = 4\pi i d_m/(A\omega\epsilon_m$), where
we introduce the quantity $\epsilon_m\equiv 1 - \omega_{J,m}^2 / \omega^2$.
Here the screened Josephson plasma frequency is defined as
$\omega_{J,m}^2 = 4 \pi \Phi_0^{-1} \epsilon_{\infty}^{-1} d_m J_m$,
and $d_m$ is the distance between neighboring planes. 
The expression for the macroscopic dielectric function $\epsilon$
follows immediately after identifying
$Z A / D$ with the complex conductivity $1/\sigma$, and by
using $\epsilon=4\pi i \sigma/\omega$. This model
can be easily extended to take into account quasi-particle currents
by shunting each $Z_m$ with a resistor in series with
an induction, resulting in an expression for $\epsilon$ analogous
to the Gorter-Casimir two-fluid model, with a normal-fluid
component $\sigma_n=\sigma_0/(1-i\omega\tau)$.
Polarization of the ions and
transverse optical phonons can be incorporated by adding separate 
oscillators to the expression for $\epsilon$. 
The expression for $\epsilon$ becomes 
\begin{equation}
 \epsilon = \left(\sum_{m} \frac{w_m}{\epsilon_{m}}\right)^{-1}
 \label{epstot}
\end{equation}
where $w_m=d_m/D$ is a weighting factor proportional to
the distance between the layers. The total dielectric function is
a sum over response functions of sets of two coupled planes
for which we obtain
\begin{equation}
 \epsilon_{m} = \epsilon_{\infty}
        \left(1 - \frac{\omega_{J,m}^2} {\omega^2}\right) 
                + \frac{4\pi i}{\omega} \sigma_n
 \label{epsm}
\end{equation}
\section{BILAYER SUPERCONDUCTORS}
Let us now consider a crystal which has two superconducting layers per unit
cell. We denote the Josephson coupling constants as I and K, the
distances between the planes as $d_I$ and $d_K$, and the screened
Josephson plasma frequencies as $\omega_I$ and $\omega_K$. 
Using Eq. (3), and approximating the effect of ionic and vibrational
screening with a constant $\epsilon_{\infty}$, the dielectric function is
\begin{equation}
 \epsilon =
 \frac{\epsilon_{\infty}(\omega^2-\omega_I^2)(\omega^2-\omega_K^2)}
        {\omega^2(\omega^2-\omega_T^2)} 
 \label{epsbi}
\end{equation}
where we introduced 
$\omega_T^2 = (d_I\omega_K^2  + d_K \omega_I^2)/(d_I+d_K)$.
Interestingly, there are now {\em two} longitudinal plasmons.
Also the dielectric function has a pole at $\omega_T$, which 
corresponds to a {\em transverse}
optical plasma oscillation, with oscillator-strength 
$S_{T}=\epsilon_{\infty}
(1-\omega_I^2/\omega_T^2)(1-\omega_K^2/\omega_T^2)$. 
\begin{figure}[th]
           \begin{center} 
           \leavevmode
           \hbox{%
            \epsfxsize=85mm
            \epsfbox{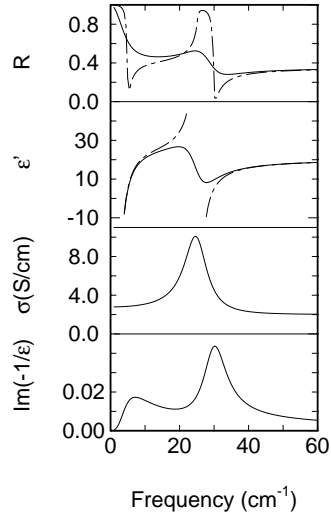}}
           \end{center} 
\caption{Reflectivity, $\mbox{Re}\epsilon$, optical conductivity and loss
function calculated for a hypothetical bi-layer superconductor,
assuming $\omega_I=5\mbox{cm}^{-1}$,$\omega_T=26\mbox{cm}^{-1}$, and
$\omega_K=30\mbox{cm}^{-1}$. The chained
curves were calculated assuming no normal-fluid component in the
conduction, the solid curves were calculated assuming a parallel 
conductance of 2 S/cm.}
\label{fig1}
\end{figure}
Photons with frequencies
within the {\em Reststrahlenband} $\omega_T<\omega<\omega_K$, and
with $\vec{E}$ polarized along $z$ can not
propagate inside the crystal and are totally reflected at the interface.
Such a reflectivity spectrum of 
the crystal-face perpendicular to the superconducting layers is displayed
in Fig. 1, along with the dispersion relation. We see, that a lifting of
degeneracy between the two Josephson frequencies $\omega_I$ and $\omega_K$
should be easily recognizable in the reflectivity curve, even if the
features in $\sigma$ are blurred due 
to parallel conduction of quasi-particles. The compound
Bi$_2$Sr$_2$CaCu$_2$O$_8$ may exhibit such behaviour.
There a resonance found at 5 cm$^{-1}$\cite{matsuda,tsui2} 
is probably the lowest of the
two Josephson plasmons. The higher plasma frequency has not yet been
observed with infrared spectroscopy down to  
$\approx$ 30 cm$^{-1}$\cite{tajima} and can be situated anywhere
between 5 and 30 cm$^{-1}$.  
\section{SINGLE LAYER SUPERCONDUCTORS}
\begin{figure}[bth]
           \begin{center} 
           \leavevmode
           \hbox{%
            \epsfxsize=85mm
            \epsfbox{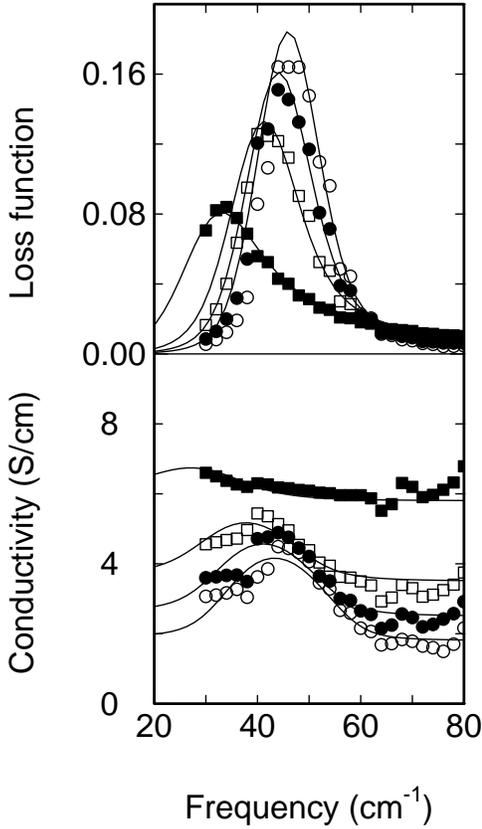}}
           \end{center} 
\caption{Experimental loss function (upper panel) and conductivity
(lower panel) of La$_{1.85}$Sr$_{0.15}$CuO$_4$ at 10 K (open circles),
15 K (close circles),  20 K (open squares), and 27 K (closed squares).
The solid lines are a fit with the same relative gaussian distribution with 
a FWHM which is $20 \%$ of the centerfrequency.}  
\label{fig2}
\end{figure}
We now apply Eq. \ref{epstot} to a stack of one superconducting
layer per unit cell, but with a distribution of the interlayer coupling
around a center value.
As was the case for the ordered superlattices discussed above, the
material is effectively homogeneous {\em vis a vis}  the
optical response of a lightbeam reflecting on the $ac$-face of the
crystal, as long as the length scale of variations of $\epsilon_m$ is
much smaller than the wavelength of the electromagnetic radiation. Let us
now consider the case where a spread exists of the Josephson frequencies
around a center frequency. 
The source of such could be the local variations
in the potential barrier, {\em e.g.} due to random substition of La$^{3+}$ 
with Sr$^{2+}$ ions in the tunneling barrier between 
the superconducting CuO$_2$ layers in La$_{2-x}$Sr$_{x}$CuO$_{4}$. 
The expression for $\epsilon$ then becomes
\begin{equation}
 \frac{1}{\epsilon} = \int d\omega_J \frac{\rho(\omega_J)\omega^2}
          {\epsilon_{\infty}(\omega^2 - \omega_J^2) 
                + 4\pi i \omega \sigma_n  }
\label{epsrho}
\end{equation}
where $\rho(\omega_J)$ is the normalized distribution function of the
screened Josephson plasma frequencies. 
The distribution $\rho(\omega_J)$
should decay sufficiently rapidly for $\omega_J \rightarrow 0$.
Physically this corresponds to the condition that the weakest Josephson
junctions in the stack of layers should still have a finite critical
current (or $\omega_J$ value). 
\\
A peak in the optical conductivity which coincides with the
longitudinal Josephson plasma has been reported by at least four
different groups\cite{uchida,gerrits,kim,basov}, but has to our knowledge
not been explained in a satisfactory way. In Fig. 2 we display both
the experimental loss function and the conductivity for the region
near the Josephson plasma frequency. These data were adopted from the
paper by Kim {\em et al.}. So far we had the data of
Gerrits {\em et al.} and Kim {\em et al.} at our disposal, and checked 
the validity of our approach for all temperatures reported for these
two sets of data. The agreement with the formalism outlined above is
very satisfactory at all temperatures. To demonstrate this result we
display in the top panel of Fig. 2 the loss function at 10 K of
Kim {\em et al.}. In the lower panel the optical conductivity is displayed.
Clearly both the loss function and $\sigma$ peak at the same frequency of
48 cm$^{-1}$. If we assume that the 'intrinsic' value of the $c$-axis
conductivity is 1.8 S/cm in this case (dashed line in the lower panel),  
and assuming that there is a single value of $\omega_J$ of 48 cm$^{-1}$,
we obtain the dashed curve for the loss function in the top panel. The
finite width in this case is due to the presence of a finite background
conductivity, and is given by $2/\tau=8\pi\sigma_n/\epsilon_{\infty}$. With
$\sigma_n=1.8$ S/cm and $\epsilon_{\infty}\approx 22$ we obtain
$c/\tau= 2\pi \cdot 10$ cm$^{-1}$. In other
words: Experimentally the broadening of the plasmon-peak can not 
be explained as the usual lifetime effect. If we now assume 
that in addition $\omega_J$ has a Gaussian
distribution, we obtain the solid and chained curves in Fig. 2
for the loss function and the conductivity using Eq. \ref{epsrho} 
with a FWHM of 11 and 17 cm$^{-1}$ respectively.
Clearly the choice of FWHM which gives good agreement for the
loss function, also explains the anomalous peak in the conductivity both
qualitatively and quantitatively. 
\newpage
\section{CONCLUSIONS}
We discussed the possible existance of transverse optical Josephson plasmons in
layered superconductors with 2 or more layers per unit cell. A related
-but not similar- effect effect may have already been observed in
La$_{2-x}$Sr$_{x}$CuO$_4$ due to variations in the interlayer Josephson 
frequency, possibly 
due to the random substitution of La with Sr. It should be
possible to design and grow artificial superconducting multilayers with 
optical properties taylored to specific needs in the THz range. 
Because they provide an effective channel for coupling electromagnetic
radiation to the current perpendicular to the superconducting layers, 
transverse optical Josephson plasmons may become useful for the 
detection and emission of far-infrared radiation in the THz
range.
%
%

\end{document}